# USING TEVATRON MAGNETS FOR HE-LHC OR NEW RING IN LHC TUNNEL*

Henryk Piekarz #, FNAL, Batavia, IL 60510, U.S.A

*Abstract*

Two injector accelerator options for HE-LHC of $p^+$ - $p^+$ collisions at 33 TeV cms energy are briefly outlined. One option is based on the Super-SPS (S-SPS) [1] accelerator in the SPS tunnel, and the other one is based on the LER (Low-Energy-Ring) [2] accelerator in the LHC tunnel. Expectations of performance of the main arc accelerator magnets considered for the construction of the S-SPS and of the LER accelerators are used to tentatively devise some selected properties of these accelerators as potential injectors to HE-LHC.

## EXPECTED QUALITIES OF INJECTOR TO HE-LHC

Injector accelerator should transfer beam to a higher level accelerator with minimal beam losses. This is especially important for the HE-LHC where the scattered injected beam of energy in the TeV range can easily produce radiation levels not only causing quench but possibly damaging the magnets. In addition, the operations of the injector accelerator should be very robust minimizing in this way potentially lost time for the physics program with HE-LHC.

It is also important that the injector accelerator has the ability to pre-condition the injected beam in order to help optimize performance of the HE-LHC. One of the most important beam improvement options is a batch slip-stacking followed by bunch coalescing which may lead to as much as doubling the proton intensity in the bunch and as a result allow an increase of the HE-LHC luminosity by up to a factor of 4.

Finally, as the cost of HE-LHC accelerator construction and operations is expected to be very high the injector construction and operation cost should constitute only a fraction of the HE-LHC design.

## S-SPS INJECTOR CONCEPT

The arrangement of the S-SPS accelerator as injector to the HE-LHC is shown in Fig. 1. The beam batches from the pre-injector chain are first injected into the SPS, accelerated to 150 GeV, and then transferred to the S-SPS. The S-SPS accelerator is built in the SPS tunnel, so it can fully contain the SPS batch. The S-SPS accelerates beam to 1 TeV [1], or 1.3 TeV [3], and then extracts it to TI2 and TI8 beam transfer lines connecting the S-SPS with the HE-LHC. This procedure is repeated 24 times to fill both HE-LHC rings. During beam stacking the S-SPS beam passes through the HE-LHC detector's beam pipe.

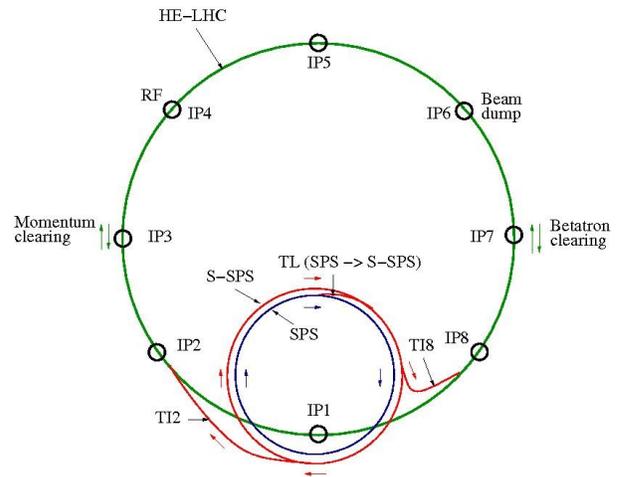

Fig. 1: S-SPS accelerator as injector to HE-LHC

The key element of the S-SPS injector proposal in [1] is that its cycle matched to the SPS eliminating the dead time incurred with the use of the S-SPS as a second stage accelerator. With the SPS beam energy set to 150 GeV its total cycle is 10.8 s. The S-SPS main arc magnet field has to be 4.5 T for 1 TeV beam and 5.9 T for 1.3 TeV one. In order to match the 10.8 s SPS cycle the ramping rate of the S-SPS magnets would have to be 1 T/s and 1.3 T/s for 1 TeV and 1.3 TeV beams, respectively. This would lead to the stacking time of 24 S-SPS beam batches in HE-LHC rings to be 4.4 minutes., as at present. As the S-SPS is also planned for the use in the fixed target experiments extending its cycle length beyond that of the SPS would cut into the benefit from the increased energy.

The increased beam energy of the S-SPS requires new construction of the TI2 and TI8 beam transfer lines to the HE-LHC using the superconducting magnets of 4 T and 5.2 T for 1 TeV and 1.3 TeV beams, respectively. The total new beam line construction for the S-SPS option is 12500 m, with 6900 m for the S-SPS ring and 5600 m for the TI2 and TI8 transfer lines.

## LER INJECTOR CONCEPT

The LER injector is a dual beam synchrotron of 1.65 TeV energy per beam placed in the LHC tunnel. The beam batches from the SPS are stacked in two LER rings and circulate in the clock-wise and counter-clock directions. As the LER rings are of the same length as the HE-LHC the LER beam batches length matches exactly those of the HE-LHC. This allows correct and improve the future HE-LHC beam batch at the LER energy. Both LER beam batches are transferred to the HE-LHC rings simultaneously using a single injection mode assuring in this way proper beam power balance in the HE-LHC two-bore magnets. The LER beam stacking time is 7.4 minutes as determined by the current SPS cycle with 24 injections (18.5 s × 24 = 444 s). The LER accelerator can work with the existing or the new pre-SPS injector chain. The SPS beam energy at the injection to the LER is 450 GeV. The LER accelerates beams to 1.65 TeV, or 10% of the HE-LHC top energy. There are two options for arranging the LER accelerator as an injector to the HE-LHC. The first option, shown in Fig. 2, allows the LER beams to bypass the detectors and the second one, shown in Fig. 3, requires the LER beams to pass through the detectors beam pipe.

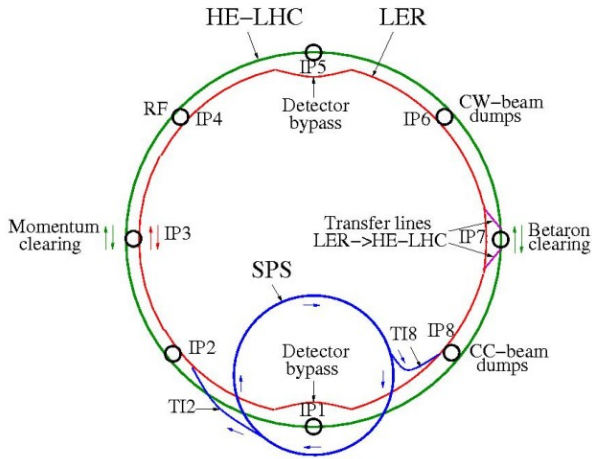

Fig. 2: LER injector Option 1 with LER beam bypassing detectors at IP1 and IP5 intersection points

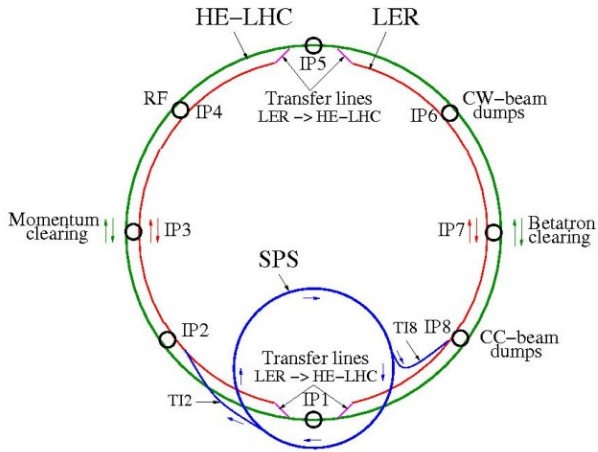

Fig. 3: LER injector Option 2 with LER beams passing the detector beam pipe at IP1 and IP5 intersection points

The advantage of the first option is that it fully secures safety of the HE-LHC during the SPS beam stacking operations in the LER. The disadvantage of this option is that construction of at least 2 x 1000 m of a new tunnel is required with the 8 T magnets used for these beam lines. For the Option 2, however, allowing the LER beams to pass through the IP1 and the IP5 intersections constitutes some risk for the detectors. For the LER Option 1 the two-beam transfer into the HE-LHC rings is enforced by two sets of kicker magnet strings located at IP7. For the LER Option 2 the two-beam transfer to the HE-LHC is enforced with total of four sets of fast switcher-magnet strings located on both sides of IP1 and IP5 intersections. The detector bypass lines in the LER Option 1 and the transfer lines in the LER Option 2 constitute an integral part of the LER synchrotron. The LER accelerator can share RF system with HE-LHC but it can also have its own installed in e.g. IP3 or IP7. For the LER Option 1 the RF system can also be placed in one of the detector bypass lines. In all cases a local expansion of the tunnel is required. The beam line construction for LER Option 1 is 26700 m long including 2000 m for the detector bypass lines and 200 m for kicker magnet strings. The total beam line construction for the LER Option 2 is 26300 m with 25904 m for the LER and 416 m for the 4 switcher magnet strings.

## EXPECTED PROPERTIES OF S-SPS MAIN ARC MAGNET

There are three crucial elements of superconducting magnet performance: (1) stability of operations (quench prevention), (2) cryogenic power loss during fast-cycling operations, and (3) overall cryogenic and electrical power demand. It is assumed [1, 3] that the S-SPS injector will use the SIS300 type magnets of the FAIR accelerator [4]. The SIS300 magnetic design [5, 6, 7] calls for a 2.75 m long dipole of $B_{max}$ = 6 T with a 50 mm gap and the d$B$/d$t$ ramping rate of 1 T/s. At present the actual tests are available for 1 m long model SIS200 dipole of $B_{max}$ = 4 T [8], and the power loss simulations for 2.6 m long SIS300 dipole of $B_{max}$ = 6 T. We extrapolate data to match the simulations (Fig. 4), and use the points at 4.5 T and 6 T to estimate the SIS300 magnet power loss at 4.5 T and 6 T for the 1 TeV and 1.3 TeV S-SPS, respectively.

Assuming the SIS300 magnet trapezoid shape of the ramping cycle 4.5 s + 1.5 s + 4.5 s = 10.5 s for the S-SPS magnet at both 1 TeV and 1.3 TeV we estimate the power loss to be 10 W/m and 15 W/m for 4.5 T and 6 T magnets, respectively. Consequently, for the 6900 m long S-SPS magnet ring of 78% filling factor the projected cryogenic power loss is 54 kW and 80 kW for operations with 1 TeV and 1.3 TeV beams, respectively.

Stability of the S-SPS accelerator operation is dependent on, among other things, the temperature margin of the superconducting magnet cable. It was analyzed in [7] that the temperature margin for a 2.6 m long SIS300 magnet operating with field cycle $B_{min}$ = 0.48 T, $B_{max}$ = 6 T, $dB/dt$ = 1 T/s, in a trapezoid time cycle 5.52-11-5.52-0 s would be no larger than 0.5 K with 40 g/s liquid helium flow. For the 6-m-long S-SPS magnet, the temperature margin will likely be even lower than 0.5 K due to the much diminished cooling efficiency in the longer cables. Consequently, one may expect the S-SPS magnet to be strongly prone to quenching and other instabilities.

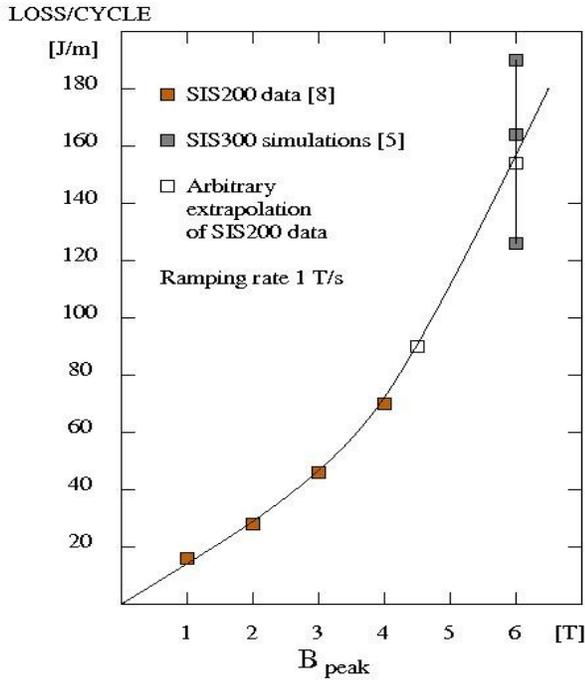

Fig. 4: SIS200 power loss data [8] for fields of 1-4 T at 1 T/s ramping rate, and extrapolation to simulations of SIS300 at 6 T

The 1 TeV S-SPS, but with ramping rate well below 1 T/s, may be a more practical solution for the large scale accelerator such as the S-SPS assuming that the lower ramping rate will indeed widen operational temperature margin. With the S-SPS as injector the stacking time in the HE-LHC rings ranges from 4.3 minutes for 1 TeV beam to 5.2 minutes for 1.3 TeV beam with the ramping rate of 1 T/s .

The electric power required for the cryogenic support (estimated using Carnot factor 70, Carnot efficiency factor 3.6 and the over-capacity factor 1.3) is 14 (17) MW for 1 (1.3) TeV S-SPS options. The ramping power of 230 kVA for the FAIR magnet scales-up to 375 (500) kVA for a 6 m long S-SPS magnet at B-fields of 4.5 T (6 T), respectively. The required ramping power for the S-SPS accelerator is then 6900 m × 0.78/6 m × 375 (500) kVA = 390 (518) MVA.

## EXPECTED PROPERTIES OF LER MAIN ARC MAGNET

A sketch of the proposed LER main arc magnet design [2] is shown in Fig. 5. This design is a scaled-down (in field) version of the VLHC Stage 1 combined function dipole [9, 10, 11]. This is a super-ferric magnet powered with a single-turn superconducting cable made of NbTi strands cooled at 4.2 K. The drive conductor with its cryostat is in the center of the magnet yoke. The return conductor is inside the cryostat pipe which supports magnetic core and houses liquid helium distribution lines for the LER accelerator. The magnet position is set with 3 posts (2 in front and 1 in rear) independently adjustable in both vertical and horizontal directions. The length of the LER magnet is 14.3 m, the same as that of the HE-LHC. The LER magnetic core cross-section is 260 mm (vertical) by 230 mm (horizontal.). Two beam gaps separated by 150 mm allow for simultaneous circulation of two proton beams in the opposite directions. For the 1.65 TeV LER synchrotron the beam gaps are 30 mm (v) · 50 mm (h), $B_{max}$ = 1.76 T, $B_{inj}$ = 0.5 T and $dB_y/dx$ = 6.5 T/m. The operating current is $I_{peak}$ = 83 kA. As the entire main arc magnet string of the LER is energized using a single-turn conductor the ramping of the accelerator is performed with a single power supply. The proposed ramping time to the full field is 60 s requiring the ramping rate of 0.02 T/s.

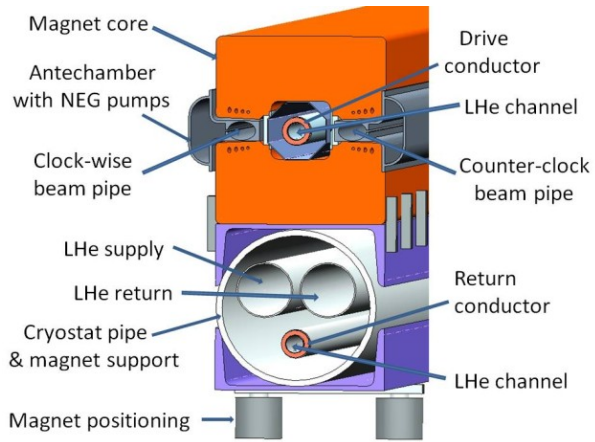

Fig. 5: LER main arc magnet position in the LHC tunnel

As in the VLHC-1, for every two dipoles there will be a set of corrector magnets consisting of horizontal and/or vertical dipole, quadrupole and sextupole magnets. The corrector magnets can be normal or superconducting. The availability of liquid helium distribution lines in HE-LHC tunnel suggests using superconducting correctors.

The stability of LER magnet cable is very high due to 2.5 K allowable temperature margin and very low static and dynamic cryogenic power losses. With 40 g/s liquid helium flow the static cryogenic heat load of the LER power cable is about 4.4 kW (scaled from the VLHC-1 design [9]). The estimated cryogenic heat load with 60 s ramping time is about 0.6 kW, leading in turn to 0.03 K temperature rise of the magnet power cable.

The total inductance of the LER accelerator ring sets the limit on the allowable cycling rates. The inductance of the LER ring (option 2, 26300 m) is about 120 mH and with 83 kA current ramping in a 60 s time period the voltage rise is 150 V. This requires the peak electrical power of 10 MVA. As the power cable can withstand much higher voltage, e.g. 1500 V, the ramping time could be shortened to e.g. 6 s with a supply of 100 MVA. The instantaneous cryogenic power loss of the LER would rise, however, to 45 kW causing the cable temperature to rise by about 2.3 K to 6.5 K, and thus approach the maximum allowable temperature of 6.9 K before quenching. Consequently, we conclude that the 6 s ramping time is not practical for the 1.65 TeV LER.

The SPS beam stacking time in two LER rings is 24 × 18.4 s (SPS super-cycle) = 7.4 minutes, and the transfer time to the HE-LHC rings equals the LER batch length of ~ 90 μs (same as the HE-LHC batch length).

It is interesting to note that the LER of 1 TeV ($B_{min}$ = 0.48 T, $B_{max}$ = 1.07 T, $dB/dt$ = 0.12 T/s) can operate with a cycle 4.5 s + 1.8 s + 4.5 s = 10.8 s thus matching the SPS cycle. The projected cryogenic power loss of ~ 8.5 kW will induce a power cable temperature rise of ~ 0.5 K. This operation reduces only slightly the allowable temperature margin from 2.5 K to 2 K. For the 1 TeV beam the LER magnet operating current is 50.5 kA and the voltage rise with 4.5 s ramping time is 670 V requiring the ramping power supply of 34 MVA. The described above LER operation can be used in the fixed target physics program, if desired.

In the LER Option 1 the accelerator sections bypassing the IP1 and IP5 intersections will use the LHC-style 8 T magnets. The power cable of these magnets will use the $Nb_3Sn$ superconductor operating at 4.5 K. With allowable temperature margin of 10 K ($T_C$ = 15 K at 8 T) it will be possible to apply 0.14 T/s ramping rate in order to reach the full field at 60 s time period. We estimate the cryogenic power loss for the 2000 m long magnet string to be about 12 kW increasing the total LER Option 1 cryogenic power to ~ 17 kW. The inductance of the 14.3 m long LHC magnet is estimated at 98.7 mH, the operating current is 11.4 kA and at the 60 s rise time there is a voltage drop of 19 V leading to about 220 kVA required ramping power. Assuming 95% magnet filling factor the two bypass beam lines will use a total of 132 magnets. The required ramping power for the bypass sections is then 29 MVA, and the total ramping power for the LER Option 1 is 39 MVA.

## S-SPS TO HE-LHC TRANSFER LINE MAGNETS

At present the SPS to the LHC TI2 and TI8 transfer line magnets are normal conducting and operate at 1.81 T field with a beam gap of 25 mm × 70 mm. For the beam energy of 1 (1.3) TeV the dipole magnetic field has to increase to 4.0 (5.2) T. This can only be achieved with superconducting magnets. One possible candidate is the Tevatron magnet (dipole is shown in Fig. 6 and quad in Fig. 7) of $B_{max}$ = 3.9 T and the radial aperture of 38 mm. As this magnet uses warm iron yoke far away from the coil the beam gap magnetic field is determined primarily by the superconductor, leading to a rather low level of higher-order multiples. Studies, however, would have to determine if such a design can be extended to higher fields. Another option is to use a cold-iron magnet, such as e.g. HERA's [12] 6 T field.

The Tevatron accelerator ring, whose circumference is comparable to the total length of TI2 and TI8 beam lines, requires 24 kW of cryogenic power at 4.2 K thus requiring about 7.9 MW of the electric power. One should expect the cryogenic power demand for the cold-iron magnets of the TI2 and TI8 beam lines to be much higher.

In summary, the S-SPS to HE-LHC beam transfer lines based on the superconducting magnets will add considerable construction and utilization costs to the HE-LHC injector chain.

The S-SPS beam would be extracted to the TI2 and TI8 lines using a combined system of kickers and septa similar to the ones used for the 450 GeV SPS beam. The kicker strength, however, will have to be considerably increased to accommodate the 1TeV or the 1.3 TeV S-SPS beams.

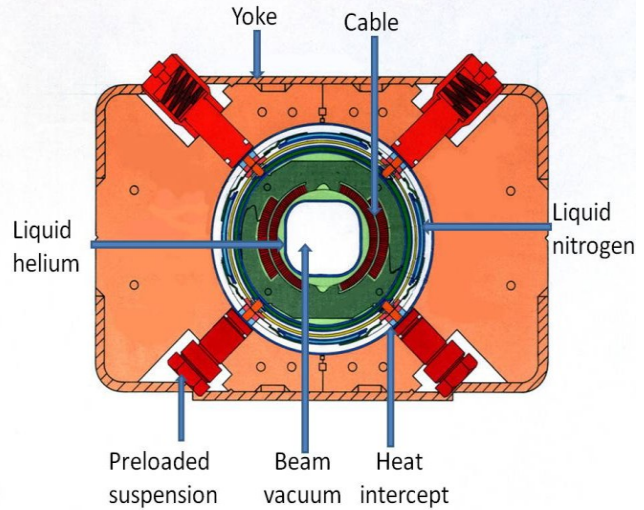

Fig. 6: Cross-section of Tevatron dipole with warm iron yoke; conductors and beam pipe are at liquid helium temperature

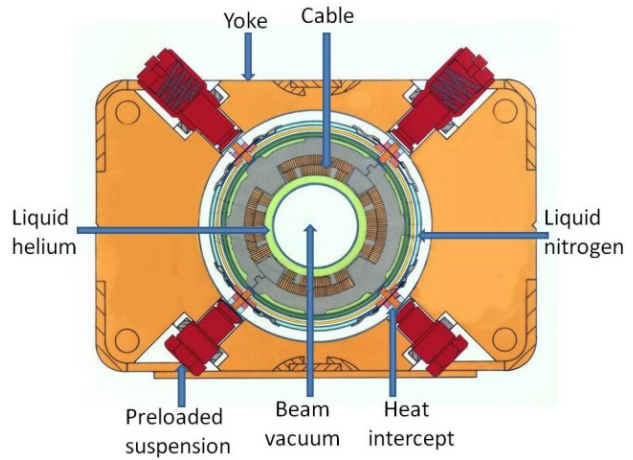

Fig. 7: Cross-section of Tevatron quadrupole with warm iron yoke; conductors and beam pipe are at liquid helium temperature

## LER TO HE-LHC TRANSFER LINE MAGNETS

### LER Injector Option 1

The simultaneous transfer of the LER beams to HE-LHC rings would take place at the IP7 area. A dual kicker magnet string of non-superconducting technology similar to the MKD beam abort system for the LHC can be used [13]. Although the 1.65 TeV LER beam energy is much lower than the 7 TeV energy of the LHC beams the beam transfer is a very challenging undertaking as a very high quality of the injected beams to the HE-LHC rings has to be preserved [13].

*LER Injector Option 2*

The LER to HE-LHC beam transfer takes place in the short straight sections around the interaction points IP1 and IP5, as described in [2]. These sections are also part of the LER Option 2 synchrotron normal operations. A dual fast-switching (3 μs time) superconducting dipole string is the key element of this beam transfer system. The principle of a fast-switching dipole is presented in [2, 14]. The HE-LHC beams separation being enlarged to 300 mm facilitates the implementation of this design. The beam separation dipoles and quads in the IP1 and IP5 sections of HE-LHC are also taking part in the LER operations. In addition, four dual-bore 8 T $Nb_3Sn$ superconducting magnets in each of the transfer lines constitute components of the LER accelerator. The estimated cryogenic power for these sections of the LER is 5 kW, and so the total LER Option 2 cryogenic power is 10 kW.

The inductance of 8 T magnets string used in the beam transfer sections of the LER Option 2 is estimated at 1.3 H thus requiring about 2.9 MVA ramping power supply for 60 s ramping time. The total required ramping power for the LER Option 2 is then about 13 MVA.

Table 1: HE-LHC beam properties at injection

| **Beam parameters** | **450 GeV** | **1 TeV** | **1.65 TeV** |
|---|---|---|---|
| RMS bunch length [cm] | 11.24 | 9.23 | 8.15 |
| RMS energy spread | $4.72 \times 10^{-4}$ | $2.58 \times 10^{-4}$ | $1.77 \times 10^{-4}$ |
| Direct space charge tune shift | $-1.54 \times 10^{-3}$ | $-3.8 \times 10^{-4}$ | $-1.58 \times 10^{-4}$ |
| Laslett tune shift | $-1.42 \times 10^{-2}$ | $-6.4 \times 10^{-3}$ | $-3.88 \times 10^{-3}$ |
| Space charge transv. impedance [MΩ/m] | $-j\,6.71$ | $-j\,3.03$ | $-j\,1.83$ |
| Space charge longit. impedance [mΩ] | $-j\,6.04$ | $-j\,1.36$ | $-j\,0.528$ |
| Microwave thresh. intensity [$N_p$/bunch] | $1.14 \times 10^{13}$ | $6.3 \times 10^{12}$ | $4.3 \times 10^{12}$ |
| Landau damping thresh. intensity [$N_p$/bunch] | $2.5 \times 10^{12}$ | $9.5 \times 10^{11}$ | $5.1 \times 10^{11}$ |
| TMCI thresh. intensity [$N_p$/bunch] | $3.0 \times 10^{12}$ | $3.7 \times 10^{12}$ | $4.2 \times 10^{12}$ |

## INJECTION ENERGY AND HE-LHC BEAM PARAMETERS

Best operation of the accelerator magnet is typically in the field range above some 10% of its top value. For the high-field type magnets the beam energy to magnetic field response is approximately linear suggesting that for the 16.5 TeV top energy the injected beam energy would be the best at 1.65 TeV, or higher. The LER accelerator can match this requirement. Beam injection energy affects beam dynamics of HE-LHC operations. The main issues are: dynamic aperture, persistent currents and snapback, instabilities, electron cloud, synchrotron radiation, and rest-gas scattering. A progression of the HE-LHC beam dynamics parameters with injection energy: 0.45 TeV, 1 TeV and 1.65 TeV is shown in Table 1 from [15]. The microwave instability threshold intensity and the Landau damping threshold intensity were found by assuming $(Z_L/n)\_eff = 0.1$ Ω, and the TMCI threshold intensity is found assuming a transverse impedance $Z_T = 3.6$ MΩ/m.

The beam size decreases with the increased energy as $1/\gamma^{1/2}$ making the physical aperture larger in *rms* units of beam size. The persistent magnet currents are reduced at higher magnetic fields (hence higher injection energy) leading to much more stable magnetic cycle. The beam instabilities due to direct space charge and beam pipe image current, etc., decrease as $1/\gamma^2$, and the rise time for the electron cloud induced instabilities increases with γ thus reducing this effect. The synchrotron radiation power increases but critical energy at beam energies up to 1.65 TeV is well below the photo-electrons work function. The emittance growth rate due to elastic scattering falls with increasing energy as $1/\gamma$ being smaller at 1.65 TeV than at 1 TeV. In summary, higher injection energy of the LER will significantly improve the long-time circulating HE-LHC beam thus minimizing its losses, reducing setup time and thus increasing the integrated luminosity.

## USING INJECTOR ACCELERATOR TO INCREASE HE-LHC LUMINOSITY

The batch slip-stacking followed by the coalescing of two bunches into a single bunch has been successfully applied at Fermilab [16]. This procedure doubles the bunch intensity, and as a result it increases instantaneous luminosity up to a factor of 4 (and so the integrated one as well). This procedure is enforced by the RF power, and for a given beam energy the higher the RF power the smaller are the beam losses. For the 450 GeV beam the particle loss is projected to be below the 5% level [2] with the RF power of 28 MV. Such an RF power (or higher) is now achievable with both normal and the superconducting RF systems. The batch slipping and bunch coalescing process would take about 11.3 s in the LER [2].

The batch slipping and bunch coalescing can also be performed in the S-SPS with the 150 GeV beams. The required RF power would be about 10 MV. This process, however, would have to be repeated 24 times for each S-

SPS batch. We estimate that the time to complete the batch slip-stacking and bunch coalescing for the SPS is about 2.9 s. The overall time for 24 batches of the S-SPS is then at least 70 s.

## DETECTOR AND HE-LHC SAFETY

The S-SPS and the LER pilot beams will be used to test the readiness of HE-LHC, the same way the SPS and LHC operate at present. The readiness of the S-SPS and LER will be tested using the SPS pilot beams. The failure of the injector before the start-up of the beam stacking in HE-LHC rings will result in time loss for the HEP physics program. The failure of the injector during the stacking process may in addition damage accelerator components. Consequently, the robustness of injector operations is of a very great importance.

The required 24 stacking operations in order to fill the HE-LHC rings with S-SPS increases the potential for aborting the stacked beams if any of the subsequent beam transfers has failed. The failed beam transfers as well as the aborted beams carry risk of damaging detectors and accelerator components. This gives an advantage to the LER where a simultaneous, single transfer of both the clock-wise and counter-clock beams will take place.

The LER magnet cable is very robust with large liquid helium channel in direct contact with the superconductor. As a result this cable can accept an instantaneous heating due to beam loss or other source of temperature rise of up to 2.7 K. In the LER Option 1 the accelerator sections for the detector bypass will use magnets based on the $Nb_3Sn$ superconductor thus likely exceeding the LER nominal operational temperature margin. In the LER Option 2 the transfer line magnets will also use $Nb_3Sn$ superconductor cable and in addition the HTS superconductor cable of the fast-switching dipoles will be set to operate with a 20 K temperature margin. The main problem with Option 2 is the necessary application of a superconducting inductor which must inject a high current into the switcher magnet cable during the 3 µs long HE-LHC beam batch gap. The failure of the inductor will result in the beam loss. A set of collimators and beam dumps as described in [2] will have to be installed in the transfer lines sections to protect the accelerator components and detectors.

In the LER Option 2 the quads and separation dipoles at the interaction points are part of the LER accelerator during the beam stacking. As the energy of the LER beam at injection and transfer to HE-LHC is low compared to the top HE-LHC energy using these magnetic components in the LER operations should be considered very safe especially since the HE-LHC quads at the IP sections will use the $Nb_3Sn$ superconducting cable.

## ARRANGEMENT OF LER AND HE-LHC MAGNETS IN LHC TUNNEL

A possible arrangement of LER and HE-LHC magnets in the LHC tunnel is shown in Fig. 8. HE-LHC magnet size was scaled-up from the LHC magnet using the cold mass diameter of 800 mm with beam separation of 300 mm, as proposed in [17]. The vertical position of HE-LHC magnet is set to 1051 mm to facilitate creation of a maximum allowable space for the transportation of another HE-LHC magnet while the one is already in place. The supporting fixtures of HE-LHC magnet are the same as for the LHC except of their increased height. The space for passing the second HE-LHC magnet is rather limited but acceptable.

The LER magnet is placed at 2123 mm height, or 1072 mm above the HE-LHC one. In working-out its location we kept all tunnel fixtures (cable trays, etc.) unchanged. Each LER magnet is supported from two columns placed between the HE-LHC magnet cryostat flanges in a way that the brackets fastening the columns to the floor do not interfere with those supporting the HE-LHC magnet, as shown in Fig. 9. The top ends of the LER columns are fastened to the tunnel ceiling providing steadiness. With this arrangement of supports both LER and HE-LHC magnets can be independently placed or removed from their accelerator rings.

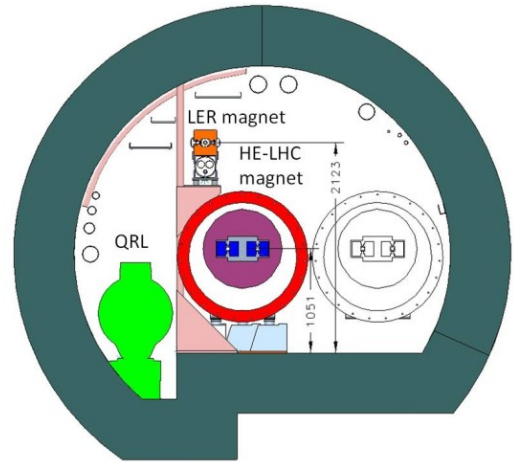

Fig. 8: Possible arrangement of LER and HE-LHC magnet rings in the LHC tunnel and position of a second HE-LHC magnet in transportation through the tunnel.

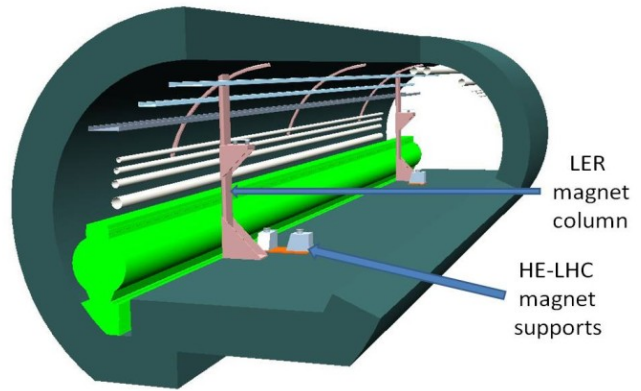

Fig. 9: Arrangement of LER magnet supporting columns relative to HE-LHC magnet supports

A perspective view of the HE-LHC and LER magnets in the LHC tunnel is shown in figure 10. The QRL cryogenic support system, all piping and cable trays are those used at present to support the LHC.

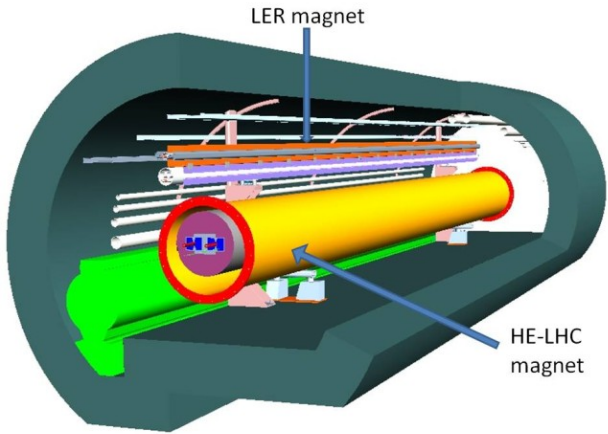

Fig. 10: Perspective view of LER and HE-LHC magnets in LHC tunnel

## S-SPS AND LER SYNCHROTRONS COST ESTIMATE

The cost of development and construction of the 20 T magnets for the HE-LHC accelerator and the cost of their supporting cryogenic and power systems will be high. Therefore, it is important to lower as much as possible the injector cost in both the construction and the utilization phases, so they will constitute only a fraction of the total HE-LHC project. For evaluation of the accelerator cost we used the total cost of a synchrotron construction rather than that of the magnet strings alone which typically may constitute only a fraction of the total synchrotron cost.

For the S-SPS accelerator the SIS300 magnets of the FAIR project are being considered. Consequently, we use the FAIR projected cost [4, 18] to estimate the cost of the S-SPS accelerator. The FAIR synchrotrons cost is a sum of 82.1 M€ for SIS100, 96.0 M€ for SIS300 and 104.4 M€ for the Common Accelerator Systems (CAS). Assuming arbitrarily that 25% of CAS cost is due to the SIS300 synchrotron our projected cost of the SIS300 accelerator is (96.0 + ¼ 104) M€ = 122 M€. The total SIS300 magnet string length in the FAIR accelerator is 454 m, and so the cost per meter of the synchrotron magnet length is 122 M€/454 m = 0.269 M€/m. Using this scaling for the S-SPS magnet string length of 6210 m (6900 m × 0.78 filling factor) the projected cost is 1490 M€.

For the cost estimate of the LER we scaled-down from the VLHC Stage 1 accelerator [9]. This cost included all accelerator subsystems: main arc magnets, correctors, RF, electric power, refrigerators, cryogenic distribution lines, accelerator controls, vacuum system and installation of all subsystems in the tunnel. With the VLHC ring length of 233 km the scaling factor for the LER is 26.6/233 = 0.12. The major material cost was corrected for the price increase of the raw materials from 2001 to 2010 using the Camden Copper and GE Commercial Finance Future of Steel price evolutions. The projected in this way LER construction cost is 170 M€. The LER Option 1 cost includes two 1000 m long beam lines bypassing detectors at IP1 and IP5 interaction points. These beam lines will use magnets based on the $Nb_3Sn$ conductor whose cost is about 4 times higher than NbTi [19]. Assuming that in the LHC-type magnet conductor constitutes 1/3 of the cost [19] we project the cost of the LER detector bypass beam lines scaling from the LHC accelerator cost (not just the magnets). The result is 170 k€/m of beam line, leading to about 326 M€ for 2000 m of the detector bypass lines. With added 50 M€ for the digging cost of a 2000 m tunnel the total cost of the LER Option 1 synchrotron is estimated at 546 M€.

## TRANSFER LINES COST ESTIMATE

For the S-SPS the new TI2 and TI8 transfer lines cost is estimated by scaling-up the 220 M€ cost of the RHIC [20] 3834 m long superconducting synchrotron. Using this scaling the estimated cost of the TI2 and TI8 beam lines is 5600/3834 × 220 = 320 M€.

For the LER Option 1 two kicker-magnet strings such as the MKD in the LHC, but with the bending power for the 1.65 TeV beam, are required to transfer beams to the HE-LHC. We estimate the cost of two 50-m-long non-superconducting kicker-magnet strings at about 10 M€.

For the LER Option 2 four superconducting magnet strings of 100 m each are required. The first 80 m length of this string uses 8 T, two-bore $Nb_3Sn$ magnets, and the remaining 20 m section uses 1.6 T HTS based fast-switching magnets. The total estimated cost of the 8 T magnets is 52 M€, and for all the fast-switching magnets we expect 48 M€, including R&D. The total estimated cost of the LER Option 2 beam transfer line sections to the HE-LHC is then 100 M€.

## SUMMARY

We presented tentatively some properties of HE-LHC injectors based on the S-SPS or the LER synchrotrons. A summary of these properties is given in Table 2. The LER injector in either of its options is superior to the S-SPS. Both LER options offer much higher injection energy and as a result much improved quality of HE-LHC beam. In addition, they allow for up to a factor of 4 increase of the HE-LHC luminosity. The LER beam stacking time is longer by about 2 minutes relative to the HE-LHC beam stacking time with the S-SPS but this is relevant only for the LER Option 2 which uses the HE-LHC ring components at IP1 and IP5 interaction points. The beam stacking time into the HE-LHC rings with the LER Option 1 is equal to the LER beam batch length of about 90 μs.

The LER Option 1 is characterized by high safety for the detectors and high reliability of its operations due to wide temperature margins of all used superconducting magnets. In addition, the LER Option 1 is independent of the HE-LHC operations, and the beam stacking in the LER rings (including bunch improvements) can be made while HE-LHC is still running the physics program. This makes the LER Option 1 injector possibly a dead-time free for the colliding beam physics.

The LER Option 2 does not require construction of new tunnels, and it will use rather short LER to HE-LHC beam transfer lines substantially minimizing the injector cost. This option, however, relies on using four strings of fast-switching magnets requiring a substantial R&D effort to make their operations secure for the detectors and for both accelerators as well.

The S-SPS synchrotron is based on the high-field, fast-cycling superconducting magnets which have not been proven yet to be applicable for a large scale synchrotron. As rather significant power losses are expected in the operations of these magnets the allowable temperature margin is very narrow suggesting a strong possibility of frequent quench occurrences and other instabilities. In view of the above the 1.3 TeV S-SPS is very unlikely to be practical. On the other hand the 1 TeV S-SPS, even if it turns out to be feasible, will not provide satisfactory improvement in the quality of the HE-LHC beams at the injection, as indicated in Table 1.

The S-SPS could also be used to double the bunch intensity before injecting its beams to the HE-LHC. This procedure, however, would have to be performed 24 times to complete the beam stacking in the HE-LHC rings. Even a very small beam loss incurred during batch slip-stacking and bunch coalescing procedures would likely raise the S-SPS magnet cable temperature making all but certain the occurrence of a quench. Consequently, there would be a high probability of long down-times for HE-LHC physics program with the implementation of procedures aimed at bunch intensity increase with the S-SPS.

Although construction and utilization cost estimates presented above are crude we can say with a reasonable confidence that the use of the S-SPS as an injector will add considerably to the HE-LHC cost. On the other hand, the LER in either of its options is consistent with the low-cost expectation for the HE-LHC injector. In addition, the required cryogenic power for all LER injector magnets constitutes only a small fraction of that for the HE-LHC, and as they are located in the same tunnel, sharing the cryogenic support system with the HE-LHC one may be possible. This option would considerably further reduce the cost of the LER injector (this potential savings was not used in the above cost estimate).

As mentioned earlier the LER Option 1 allows for safe operation of 1 TeV beams with the cycling period matching that of the SPS at 150 GeV. This operation can be used e.g. to extract beams for production of secondary beams of the fixed target physics program. In such operation all beam stacking takes place only in the SPS and the LER serves simply as an energy booster, the same way as proposed for the S-SPS. The LER super-cycle will be twice longer than that of the SPS to allow injection of the SPS beam batches into two rings of the LER. The two LER beams will be simultaneously accelerated and then extracted onto the secondary beam production targets. A comparison of some selected properties of the S-SPS and LER synchrotrons operating with 1 TeV beams for the fixed target physics program are listed in Table 3.

Table 2: Estimated properties of S-SPS and LER injectors

| Injector Properties | S-SPS | LER-1 | LER-2 |
|---|---|---|---|
| HE-LHC injection energy [TeV] | 1 (1.3) | 1.65 | 1.65 |
| Number of injections | 24 | 1 | 1 |
| Doubling bunch intensity | No | Yes | Yes |
| HE-LHC filling time [min] | 4.3 (5.2) | ~ 0 | 7.4 |
| Temperature margin [K] | 0.5 (< 0.5) | 2.5 | 2.5 |
| Quench probability | High | Very low | Low |
| Operations complexity | High | Medium | Medium |
| Synchrotron cryogenic power @ 4.2 K [kW] | 54 (80) | 17 | 10 |
| Transfer lines cryogenic power @ 4.2 K [kW] | 30 | 0 | 0 |
| Synchrotron ramping power [MVA] | 390 (500) | 39 | 13 |
| Synchrotron cost estimate [M€] | 1490 | 546 | 170 |
| Transfer line cost estimate [M€] | 320 | 10 | 100 |
| Injector cost estimate [M€] | 1810 | 556 | 270 |

Table 3: Estimated properties of S-SPS and LER Option-1 synchrotrons in application for fixed target physics program

| Synchrotron properties | S-SPS | LER-1 |
|---|---|---|
| Beam energy [TeV] | 1 | 1 |
| Number of beams | 1 | 2 |
| Operation super-cycle [s] | 10.8 | 21.6 |
| Temperature margin [K] | 0.5 | 2 |
| Cryogenic power @ 4.2 K | 54 | 27 |
| Ramping power [MVA] | 390 | 178 |

In the proposed above fixed target LER operations the cryogenic and ramping powers are increased substantially

In the proposed above LER operations the cryogenic and ramping powers are increased substantially relative to the LER Option 1 operating as an injector to the HE-LHC. This increase is mostly due to $Nb_3Sn$, 8 T magnet strings used in the construction of the HE-LHC detector bypass lines. Nevertheless, the expected four times wider temperature margin, twice lower cryogenic and ramping powers, much simplified operation control system (single power supply with single quench detection and protection systems) and much lower construction and utilization costs are all in favor of selecting the LER synchrotron rather than the S-SPS one for the fixed target physics program.

## CONCLUSIONS

We believe that the very narrow temperature margin, insufficiently high injection energy and very high cost of construction and utilization make the S-SPS synchrotron an unlikely candidate as injector to the HE-LHC. On the other hand, the 1.65 TeV LER Option 1 synchrotron with its wide temperature margin, optional doubling of the HE-LHC bunch intensity and moderate construction and utilization costs, should be considered as the primary candidate for the injector to the HE-LHC accelerator.

The LER Option 2 can be considered for the HE-LHC injector only after proving that the LER to HE-LHC beam transfer using fast-switching superconducting magnets is robust and safe for both the detectors and accelerators. We believe that the R&D effort to develop the fast-switching superconducting magnets is warranted as potential saving in the LER injector cost is high not only in the relative but more importantly in the absolute terms. In addition, this new superconducting magnet technology if successful will be very useful for other accelerator sub-systems e.g. kicker magnets, high-current dump switches, etc., as well as for the high-current superconducting cable industrial applications.

During the HE-LHC colliding beam period, the LER Option 1 accelerator can be safely used for the fixed target physics programs with the selection of the extracted beam energies from 0.45 TeV to 1 TeV, and up to 1.65 TeV, if the LER super-cycle is extended beyond the SPS one.

## ACKNOWLEDGEMENTS

It is my pleasure to thank Jamie Blowers, Steve Hays and Tanaji Sen for reading the manuscript and several valuable discussions.

## REFERENCES

[1] W. Scandale, "LHC upgrade based on high intensity, high luminosity, high energy injector chain" LHC-LUMI-05, 2005
[2] G. Ambrosio et al., "LER-LHC Injector Workshop Summary", LUMI-06 Workshop", Valencia, 2006
[3] F. Zimmerman, "LHC Beyond 2020", KEK Seminar, 2010
[4] FAIR Baseline Technical Report – *Executive Summary*, ISBN 3-9811298-0-6, 50, 2006
[5] P. Scherbakov et al., "Comparative analysis of wide aperture dipole designs for the SIS300 ring", Proc. RuPAC XIX, Dubna, 301-303, 2004
[6] I. Bogdanov et al., "Study of the quench process in fast cycling dipole for the SIS300 ring", EPAC 2004, 1744-1746, Lucerne
[7] V. Zubko et al., "Stability of fast-cycling dipole for the SIS300 ring", EPAC 2004,1756-1758, Lucerne
[8] M.N. Wilson et al., "Measured and Calculated Losses in Model Dipole for GSI Heavy Ion Synchrotron", IEEE Trans. on Applied Superconductivity, 14, 306-309, 2004
[9] *VLHC Design Study*, Fermilab-TM-2149, 2001
[10] H. Piekarz et al."A Test of a 2 T Superconducting Transmission Line Magnet System", IEEE Trans. Appl. Superconductivity 16, 342-345, 2006
[11] G. Velev et al., "Field Quality Measurements of a 2 T Transmission Line Magnet",IEEE Trans. Appl. Superconductivity, 16, 1840-1843, 2006
[12] K.-H. Mess, P. Schmüser and S. Wolf, "Superconducting Accelerator Magnets", ISBN 981 02- 2790-6, 1996
[13] B. Goddard, "LHC beam dump, injection system and other kickers", Workshop on HE- LHC, Malta, Oct. 14-16, 2010
[14] S. L. Hays, "High-Current Superconducting Inductor", Private Communication, October, 2010
[15] T. Sen, "Estimate of HE-LHC beam parameters at different injection energies", FERMILAB-TM-2478 –APC, 2010
[16] K. Seiya, et al, "Multi-batch slip-stacking in the Main Injector at Fermilab", PAC07, 742-743, 2007
[17] E. Todesco and L. Rossi, "Conceptual Design of 20 T Dipoles for Higher Energy LHC", Workshop on HE-LHC, Malta, Oct. 14-16, 2010
[18] C. Muehle, "Fast-Pulsed Superconducting magnets", HB2006, Tsukuba, 324-328, 2006
[19] E. Todesco, *private communication*, 2010
[20] M. Harrison, S. Peggs and T. Roser, "The RHIC Accelerator", Annu. Rev. Nucl. Part. Sci., 52:425-69, 2002